\documentclass{webofc}
\usepackage[varg]{txfonts}   
\begin{document}
\title{New vistas in charm production}
%
%

\author{\firstname{Antoni}
  \lastname{Szczurek}\inst{1,2}\fnsep\thanks{\email{ifj.edu.pl}} 
}

\institute{Institute of Nuclear Physics PAN, Krak\'ow, Poland
\and
           Rzesz\'ow University, Rzesz\'ow, Poland
          }

\abstract{%
We discuss some new aspects of charm production trigerred by recent
observations of the LHCb collaboration.
The LHCb collaboration measured small asymmetries in production of
$D^+ D^-$ mesons as well as $D_s^+ D_s^-$ mesons.
Is this related to initial quark/antiquark asymmetries in the proton ?
Here we discuss a scenario in which unfavored fragmentations
$q/{\bar q} \to D$ and $s/{\bar s} \to D_s$ are responsible for 
the asymmetries.
We fix the strength of such fragmentations -- transition probabilities,
by adjusting to the size of the LHCb asymmetries.
This has consequences for production of $D$ mesons in forward directions
(large $x_F$) as well as at low energies.
Large asymmetries are predicted then in these regions.
We present here some of our predictions. Consequences for high-energy
neutrino production in the atmosphere are discussed and quantified.
The production of $\Lambda_c$ baryon at the LHC is disussed.
Large deviations from the independent-parton fragmentation picture
are found.
}
\maketitle
%
\section{Introduction}
\label{intro}

It is usually assumed that the $c/{\bar c} \to D$ 
fragmentation is responible for production of charmed mesons.
In leading order $g g \to c \bar c$ is dominant partonic subprocess.
The contribution of $q {\bar q} \to c {\bar c}$ is usually much smaller.
The leading-order production of charm is by far insufficient to
describe experimental distributions of $D$ mesons in rapidity and
transverse momentum. The NLO calculation is needed to describe
experimental data. An alternative is the $k_t$-factorization approach
which gives resonable description of $D$ meson single particle
distributions \cite{Maciula:2013wg}. It allows to describe even some 
correlation observables \cite{hms2014}. Usually the Peterson fragmentation 
functions \cite{Peterson:1982ak} are used for $c {\bar c} \to D$ 
fragmentations.

\begin{figure}[!h]
\begin{minipage}{0.45\textwidth}
 \centerline{\includegraphics[width=0.9\textwidth]{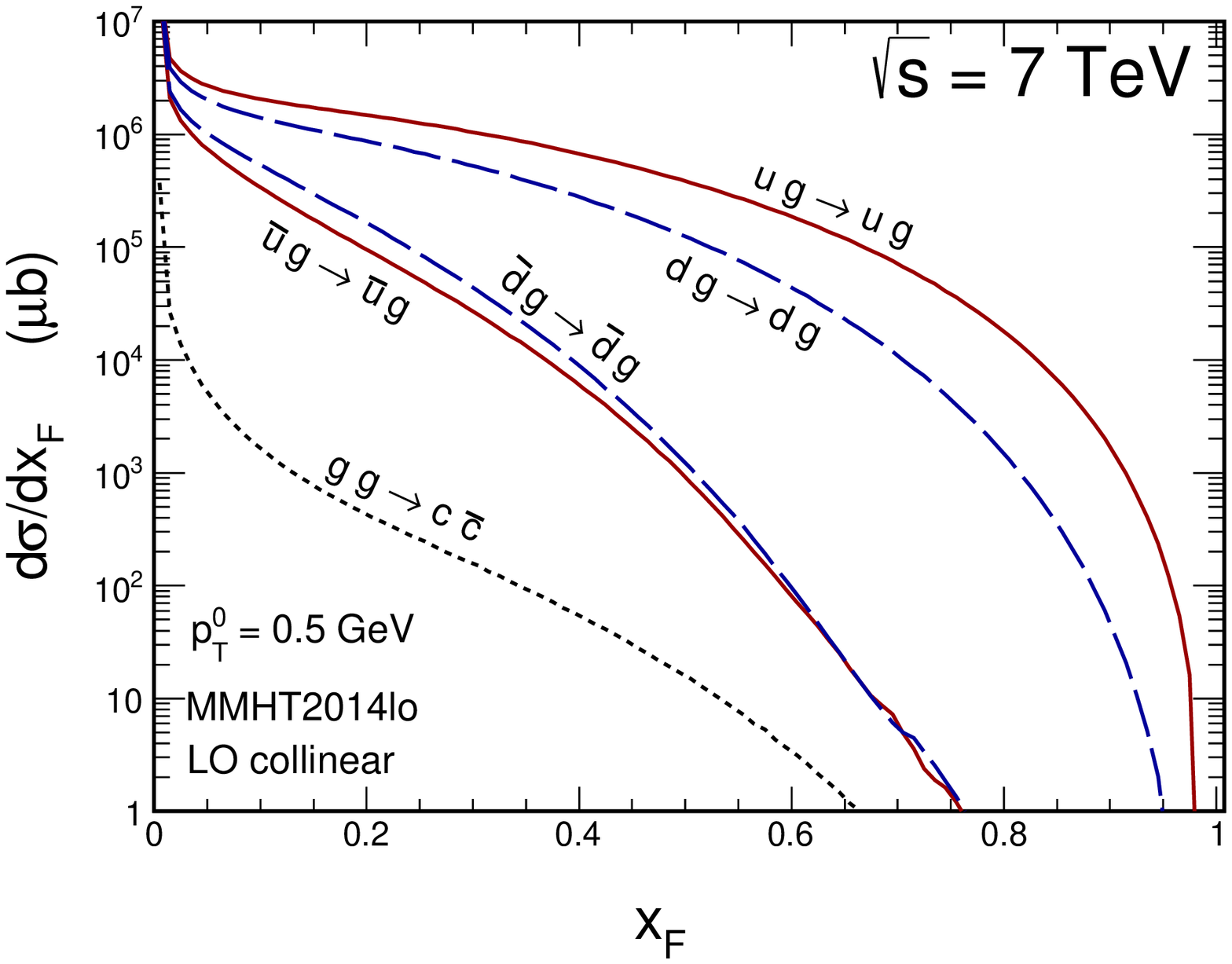}}
\end{minipage}
\hspace{0.5cm}
\begin{minipage}{0.45\textwidth}
 \centerline{\includegraphics[width=0.9\textwidth]{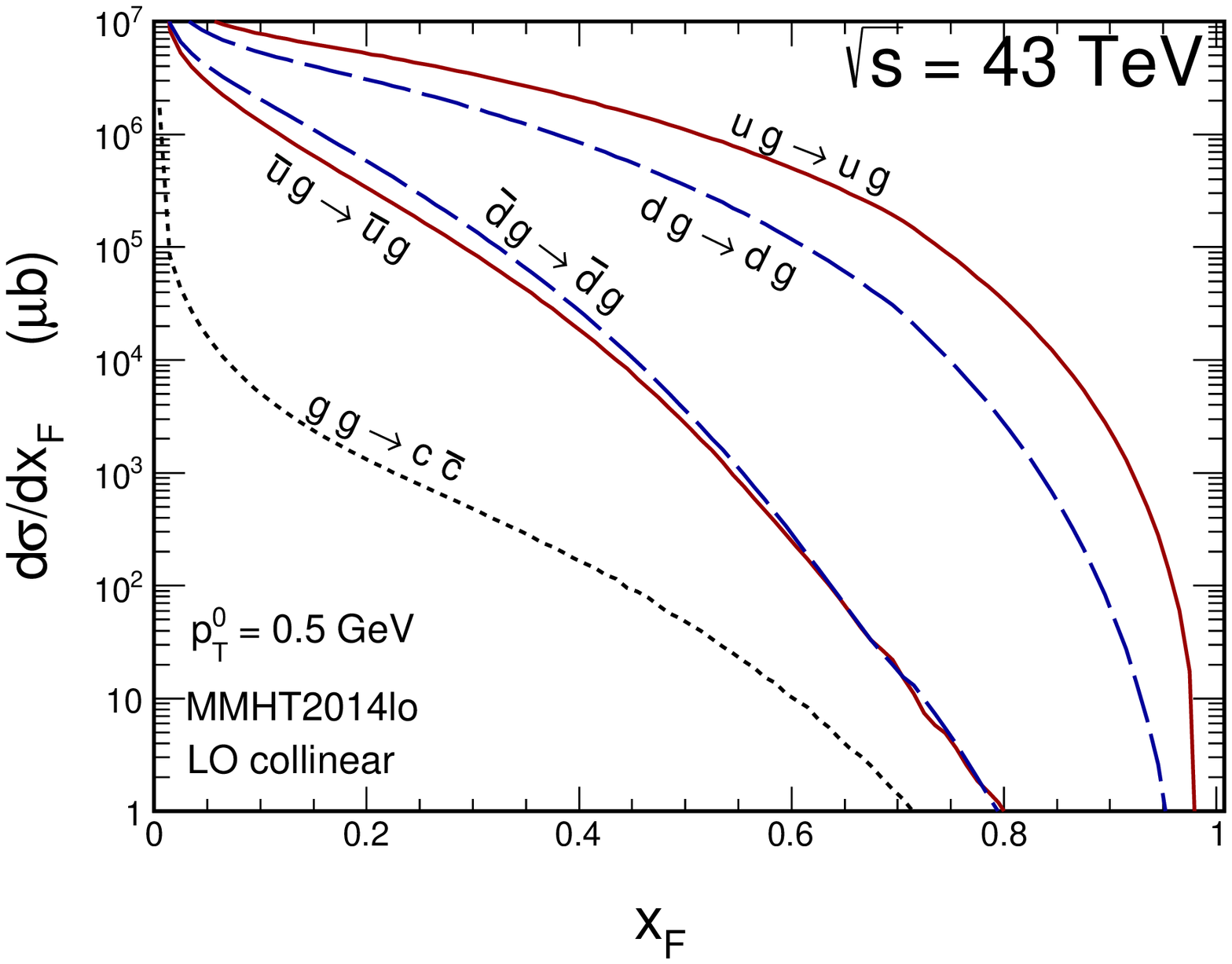}}
\end{minipage}
   \caption{
\small Quark and antiquark distributions in Feynman $x_F$ for 
$\sqrt{s} =$ 7 TeV (left panel) and $\sqrt{s} =$ 43 TeV (right panel)
corresponding to $E_{\mathrm{lab}}(p)$ = 10$^{9}$ GeV.
This calculation was performed within collinear-factorization 
approach with somewhat arbitrary regularization parameter 
$p_{T}^{0} =$ 0.5 GeV \cite{MS2018}.
 }
 \label{fig:dsig_dxf_partons}
\end{figure}

Recently the LHCb collaboration observed an intriguing asymmetries
for $D^+ D^-$ \cite{LHCb:2012fb} and $D_s^+ D_s^-$
\cite{Aaij:2018afd} production.
The question arises what is origin of such asymmetries.
In general, there can be a few reasons such as electroweak corrections,
higher-order pQCD effects. The electroweak corrections should
be important rather at large transverse momenta.
The LHCb collaboration measured the asymmetries at rather small
transverse momenta where statistics is enough to pin down the small
asymmetry effect.
In Fig.\ref{fig:dsig_dxf_partons} we show for ilustration distribution
of partons obtained in LO collinear approach.
Furthermore the distribution of light quarks and even antiquarks is 
much larger than the distribution of $c/{\bar c}$ quarks/antiquarks
produced in gluon-gluon fusion process.
The distribution of light quarks is much larger than distribution
of corresponding antiquarks.
All this suggests that a nonzero subleading fragmentation
$d \to D^-$ and ${\bar d} \to D^+$ would produce an asymmetry
when added to the dominant $c/{\bar c} \to D$ fragmentation.   
For $D_s$ meson production asymmetry the situation is more
subtle as far as subleading fragmentation is considered.
Here we have ${\bar s} \to D^+$ and $s \to D^-$ subleading
fragmentations. If $s(x) = {\bar s}(x)$ then of course
the asymmetry is zero. There are no deep reasons to assume
$s(x) = {\bar s}(x)$. Actually the nonperturbative
effects of the strange meson cloud lead to $s(x) \ne {\bar s}(x)$
(see e.g.\cite{Holtmann}).
Also some fits of parton distributions allow for different distributions
of $s$ and $\bar s$ partons \cite{Lai:2007dq}.

\section{Cross sections, production asymmetry 
and subleading fragmentations}

Let us discuss first the dominant at the LHC contribution -- the
gluon-gluon fusion. The multi-diferential cross section for $c \bar c$ 
productions can be then calculated as:
\begin{eqnarray}\label{LO_kt-factorization} 
\frac{d \sigma(p p \to c \bar c \, X)}{d y_1 d y_2 d^2p_{1,t} d^2p_{2,t}} &=&
\int \frac{d^2 k_{1,t}}{\pi} \frac{d^2 k_{2,t}}{\pi}
\frac{1}{16 \pi^2 (x_1 x_2 s)^2} \; \overline{ | {\cal M}^{\mathrm{off-shell}}_{g^* g^* \to c \bar c} |^2}
 \\  
&& \times  \; \delta^{2} \left( \vec{k}_{1,t} + \vec{k}_{2,t} 
                 - \vec{p}_{1,t} - \vec{p}_{2,t} \right) \;
{\cal F}_g(x_1,k_{1,t}^2) \; {\cal F}_g(x_2,k_{2,t}^2) \; \nonumber ,   
\end{eqnarray}
where ${\cal F}_g(x_1,k_{1,t}^2)$ and ${\cal F}_g(x_2,k_{2,t}^2)$
are the gluon uPDFs for both colliding hadrons and 
${\cal M}^{\mathrm{off-shell}}_{g^* g^* \to c \bar c}$ is the off-shell 
matrix element for the hard subprocess.
First the distribution in rapidity and transverse momentum of 
$c$ or ${\bar c}$ are obtained (inclusive cross section).
The cross section for $D$ meson can be obtained then as a convolution
of the partonic cross section for $g^* g^* \to c {\bar c}$ and 
the $c / {\bar c} \to D$ fragmentation
functions. The Peterson fragmentation function \cite{Peterson:1982ak}
with $\epsilon$ parameter adjusted to experimental data.

In the studies presented here we include also $u,\bar u, d, \bar d \to D^i$
parton fragmentation to $D$ mesons.
We include only fragmentations of quarks/antiquarks that 
are constituents of the $D$ meson.
We assume the following symmetry relation: 
\begin{equation}
D_{d \to D^-}(z) = D_{\bar d \to D^+}(z) = D^{(0)}(z) \; .
\label{ff_symmetries}
\end{equation}
Similar flavor symmetry relations hold for fragmentation 
of $u$ and $\bar u$ to $D^0$ and $\bar D^0$ mesons.\\
However $D_{q \to D^0}(z) \ne D_{q \to D^+}(z)$, which is caused
by the contributions from decays of vector $D^*$ mesons.
Furthermore we assume for doubly suppressed fragmentations:
\begin{equation}
D_{\bar u \to D^{\pm}}(z) = D_{u \to D^{\pm}}(z) = 0 \; .
\label{neglected_ff}
\end{equation}
The fragmentation functions at sufficiently large scales undergo 
DGLAP evolution equations. Since in the presented here analysis
we are interested in small transverse momenta (small scales for 
DGLAP evolution) we can just use rather the initial conditions for 
the evolution, which are for the subleading fragmentation rather 
poorly known.

We parametrize the unfavoured fragmentation functions as:
\begin{equation}
D_{q_f \to D}(z) = A_{\alpha} (1-z)^{\alpha} \; .
\label{ff_simple_parametrization}
\end{equation}
Instead of fixing the uknown $A_{\alpha}$ we will operate rather with
the fragmentation probabilities:
\begin{equation}
P_{q_f \to D} = \int dz \; A_{\alpha} \left( 1 - z \right)^{\alpha} \; .
\label{Dff_simple_parametrization}
\end{equation}
and calculate corresponding $A_{\alpha}$ for a fixed 
$P_{q \to D}$ and $\alpha$.
Therefore in our effective approach we have only two free parameters.

Another simple option we considered in \cite{MS2018} is:
\begin{equation}
D_{q_f \to D}(z) = P_{q_f \to D} \cdot D_{\mathrm{Peterson}}(1-z) \; .
\label{Peterson}
\end{equation}
Then again $P_{q_f \to D}$ would be the only free parameter.

The flavour asymmetry in production of $D$ mesons is defined as:
\begin{equation}
A_{D^+/D^-}(\xi) 
= \frac{ \frac{d \sigma_{D^-}}{d \xi}(\xi) - \frac{d \sigma_{D^+}}{d \xi}(\xi) }
       { \frac{d \sigma_{D^-}}{d \xi}(\xi) + \frac{d \sigma_{D^+}}{d \xi}(\xi) }
\; ,
\label{asymmetry_DpDm}
\end{equation}
where $\xi = x_F, y, p_T, (y,p_T)$.

For $D_s$ mesons we define the production asymmetry as:
\begin{equation}
A_{D_s^+/D_s^-}(\xi) = 
\frac{ \frac{d\sigma(D_s^+)}{d\xi}(\xi) - \frac{d\sigma(D_s^-)}{d\xi}(\xi) }
     { \frac{d\sigma(D_s^+)}{d\xi}(\xi) + \frac{d\sigma(D_s^-)}{d\xi}(\xi) }
\; .
\label{asymmetry_DspDsm}
\end{equation}

The production of $D_s$ mesons is interesting in the context of the fact
that $D_s$ mesons are the main source of $\tau$-neutrinos:
\begin{eqnarray}
&&D_s^+ \to \tau^+ + \nu_{\tau}   \; .          \\
&&D_s^- \to \tau^- + \overline{\nu}_{\tau}   
\end{eqnarray}
and in addition:
\begin{eqnarray}
&&\tau^+ \to {\bar \nu}_{\tau}+X \; ,   \\
&&\tau^- \to \nu_{\tau}+X \; .
\end{eqnarray}
Both emissions should be included in final evalution of $\tau$-(anti)neutrinos.
 
Finally in this presentation we consider production of $\Lambda_c$ baryons.
Whether the independent parton fragmentation works for $\Lambda_c$ baryons
was discussed in \cite{MS_Lambdac}.
In such an approach the cross section can be written as:
\begin{equation}
\frac{d \sigma(pp \rightarrow h X)}{d y_h d^2 p_{t,h}} \approx
\int_0^1 \frac{dz}{z^2} D_{c \to h}(z)
\frac{d \sigma(pp \rightarrow c X)}{d y_c d^2 p_{t,c}}
\Bigg\vert_{y_c = y_h \atop p_{t,c} = p_{t,h}/z} \;,
\label{Q_to_h}
\end{equation}
where $p_{t,c} = \frac{p_{t,h}}{z}$ and $z$ is the fraction of
longitudinal momentum of charm quark $c$ carried by a hadron 
$h =D, \Lambda_c$.
A typical approximation in this formalism assumes $y_h = y_c$.

\section{Results}

In this section we will show our results for (anti)neutrino
production, cross sections for $D^+ D^-$ production and 
$D^+ D^-$ and $D_s^+ D_s^-$ asymmetries as well as a possible
consequences for $\tau$ (anti)neutrino production and finally for 
$\Lambda_c$ baryon production.

\subsection{Neutrino production in the atmosphere}

\begin{figure}[t]
\begin{center}
\includegraphics[scale=0.3]{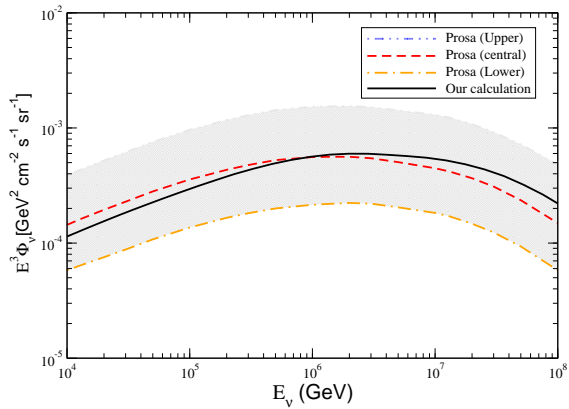}
\caption{Comparison of our predictions for the prompt neutrino flux
  and the Prosa results.}
\label{fig:neutrino_flux_vs_PROSA}
\end{center}
\end{figure}

We start from showing our best (optimal) result for neutrino flux relevant
for the IceCube experiment. In Fig.\ref{fig:neutrino_flux_vs_PROSA}
we show our predictions obtained for calculating cross section
in the $k_t$-factorization approach with the KMR unintegrated
gluon distributions. Such an approach effectively includes higher-order
corrections as was discussed in the literature.
Our result well coincides with the PROSA
results within their uncertainty band.

The flux here was calculated within the $Z$-moment method \cite{GMPS2017}. 
In such a  calculation $\frac{d \sigma}{d x_F}(x_F,\sqrt{s})$ for
production of $D$ mesons is a crucial input.

\begin{figure}[t]
\begin{center}
\includegraphics[scale=0.3]{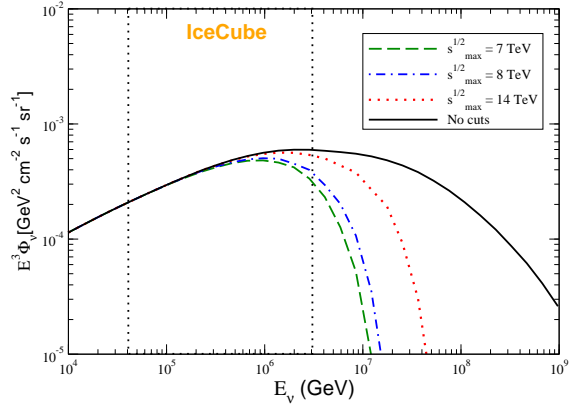}
\caption{Impact of different cuts 
on the maximal  center-of-mass $pp$ collision energy 
for the prompt neutrino flux.}
\label{fig:energycut}
\end{center}
\end{figure}

Which energies of proton-proton scattering are responsible for 
the production of high-energy neutrinos at IceCube?
In Fig.\ref{fig:energycut} we show how the upper cut on center-of-mass
energy influences the flux of high-energy neutrinos in the atmosphere.
For energies $E_{\nu} >$ 10$^8$ GeV, the collision energies larger than 
those measured at the LHC enter the calculation. So predictions are
based on extrapolation to unexplored yet region.

\begin{figure}[t]
\begin{center}
\includegraphics[scale=0.3]{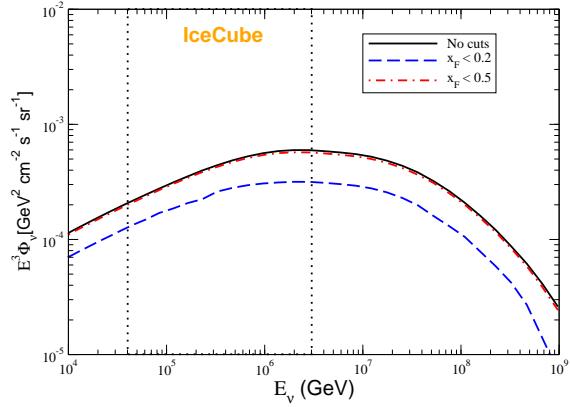}
\caption{The effect of $x_F$ cuts on the prompt neutrino flux.}
\label{fig:xf}
\end{center}
\end{figure}

What are typical Feynman $x_F$ values responsible for production
of high-energy neutrinos is illustrated in Fig.\ref{fig:xf}.
Rather large values are important. Such a region is unfortunately not
covered by the LHC detectors. Even (often called) forward LHCb detector 
is limited to $x_F <$ 0.1.

\begin{figure}[t]
\begin{center}
\includegraphics[scale=0.3]{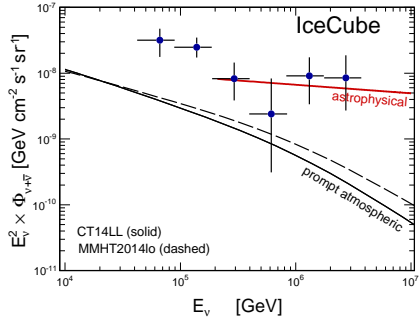}
\caption{Comparison of predictions obtained with the CT14 and MMHT PDFs
  for the prompt neutrino flux.
The data points are taken from IceCube analysis \cite{IceCube_fluxlimit}. 
For comparison, a fit for the astrophysical contribution, proposed 
in \cite{IceCube_fluxlimit} is presented as well.}
\label{fig:IceCube-data}
\end{center}
\end{figure}

In Fig.\ref{fig:IceCube-data} we show our predictions for the flux
of high-energy neutrinos. This result was obtained within
$k_t$-factorization approach. Clearly such a calculation cannot describe
the measured flux of neutrinos. No subleading fragmentations were
included here. There seems to be arguments that at least part
of the missing yield is of astrophysical origin \cite{IceCube_Science}.
Can the subleading fragmentation play a role in this context ?

\subsection{LHCb asymmetries}

\begin{figure}[!h]
\begin{minipage}{0.42\textwidth}
  \centerline{\includegraphics[width=1.0\textwidth]{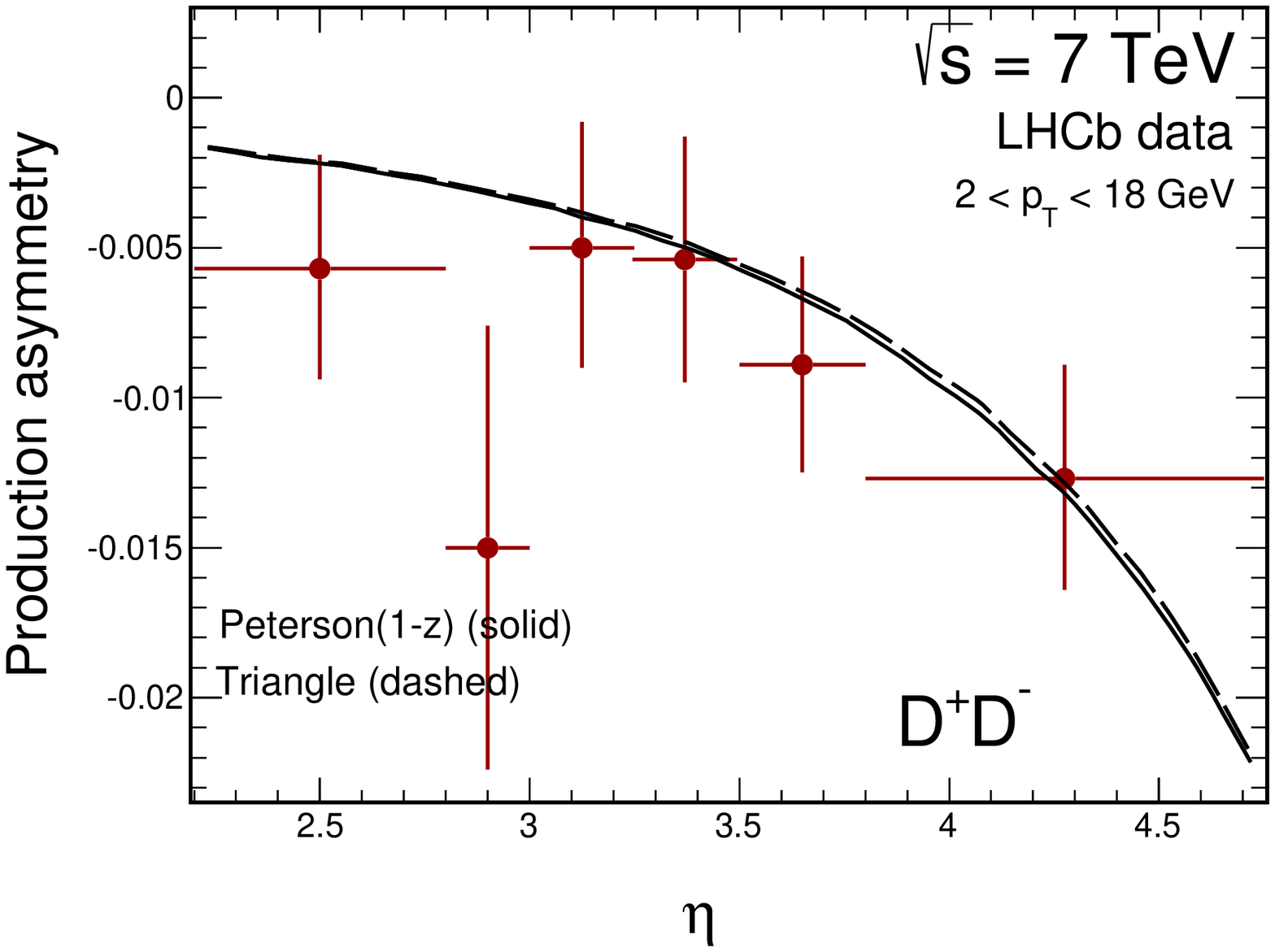}}
\end{minipage}
\hspace{0.5cm}
\begin{minipage}{0.42\textwidth}
  \centerline{\includegraphics[width=1.0\textwidth]{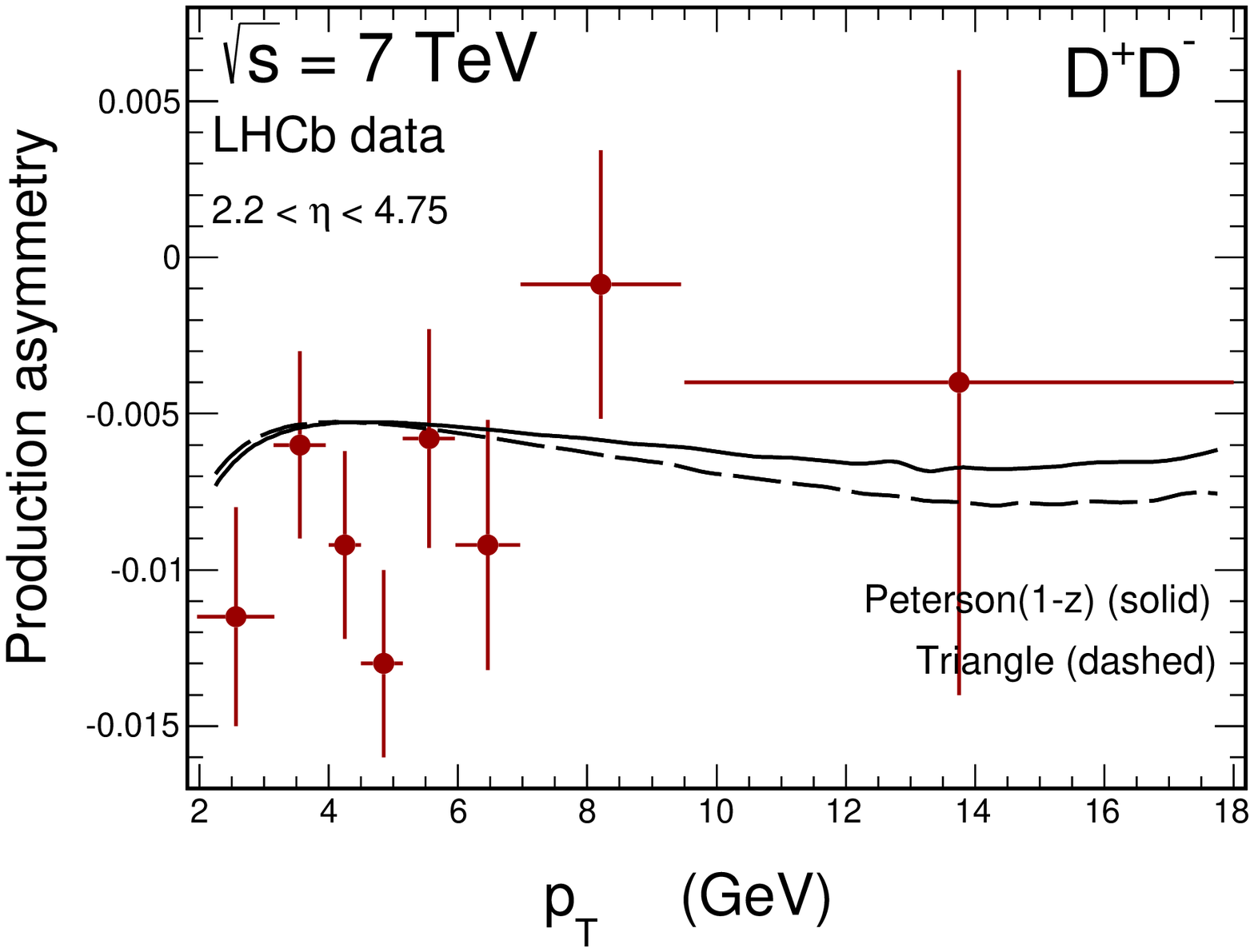}}
\end{minipage}
  \caption{
\small $A_{D^+/D^-}$ production asymmetry measured by the LHCb
collaboration at $\sqrt{s}= 7$ TeV as
a function of $D$ meson pseudorapidity (left panel) 
and $D$ meson transverse momentum (right panel). 
}
\label{fig:LHCb_asymmetry_charged}
\end{figure}

The $D^+ D^-$ asymmetries obtained by us are shown in 
Fig.\ref{fig:LHCb_asymmetry_charged} for $\sqrt{s}$ = 7 TeV. 
Only one parameter, the quark/antiquark fragmentation probability, 
was adjusted to the LHCb data.
In Ref.\cite{MS2018} we presented also our predictions for 
$\sqrt{s}$ = 13 TeV.

\begin{figure}[!h]
\begin{minipage}{0.3\textwidth}
 \centerline{\includegraphics[width=1.0\textwidth]{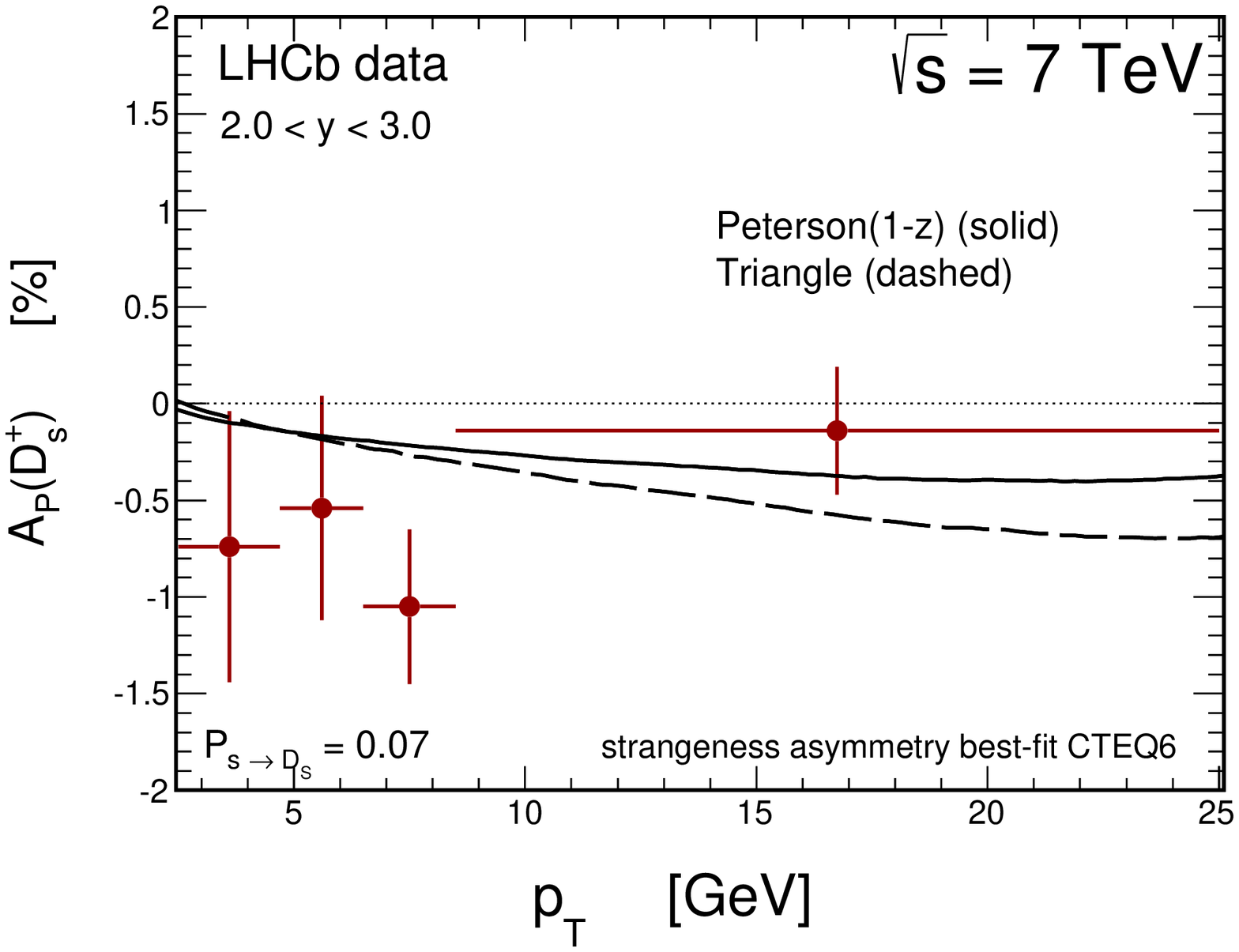}}
\end{minipage}
\hspace{0.2cm}
\begin{minipage}{0.3\textwidth}
 \centerline{\includegraphics[width=1.0\textwidth]{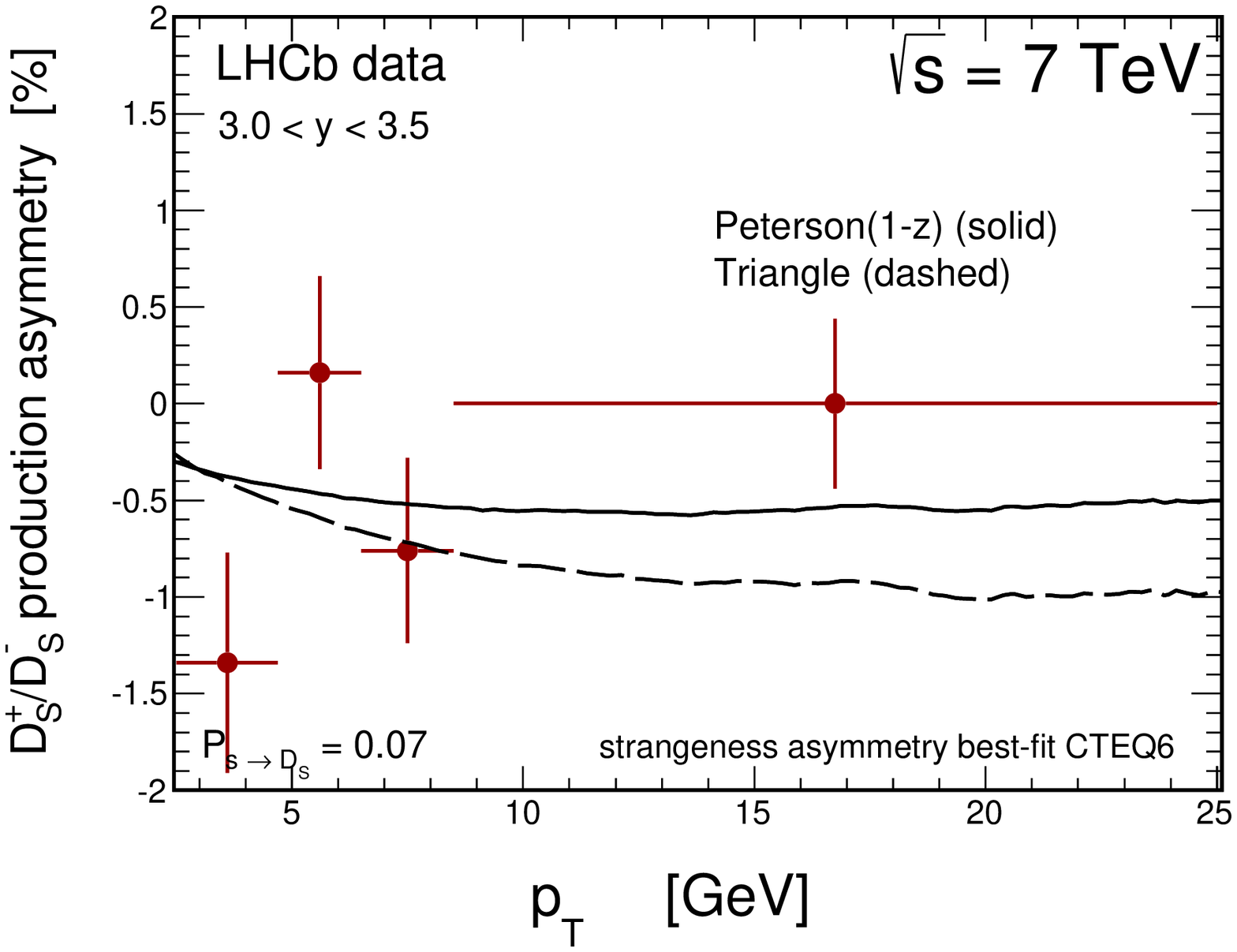}}
\end{minipage}
\hspace{0.2cm}
\begin{minipage}{0.3\textwidth}
 \centerline{\includegraphics[width=1.0\textwidth]{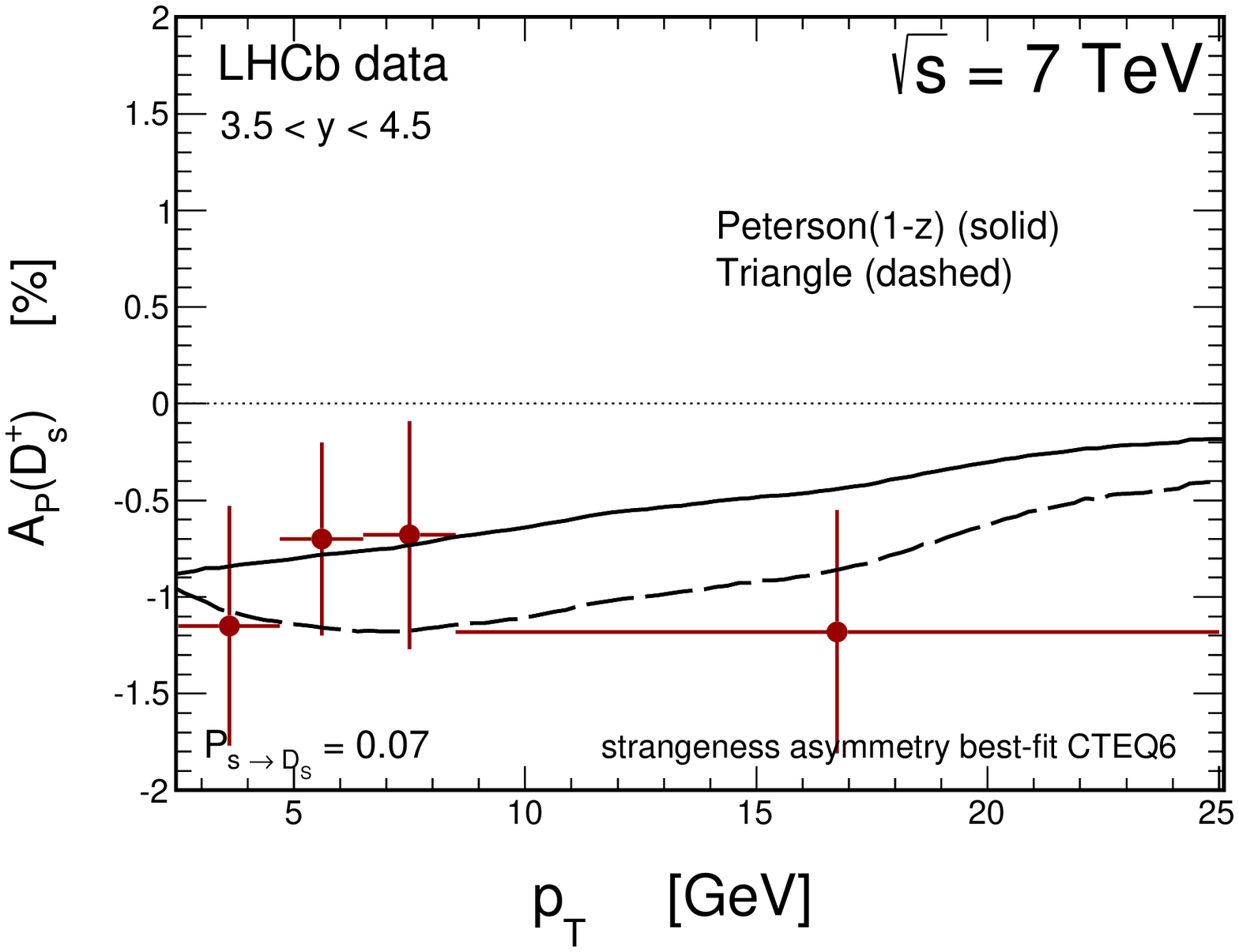}}
\end{minipage}
   \caption{$D_s^+ / D_s^-$ asymmetry obtained by us together
with the LHCb collaboration for $\sqrt{s}$ = 7 TeV. 
The CTEQ6.5 parton distributions are used in this calculation.
 }
 \label{fig:Ds_asymmetry}
\end{figure}

Similar asymmetry for the $D_s^+ D_s^-$ production is shown in
Fig.\ref{fig:Ds_asymmetry}.
Here the error bars are even larger than for the $D^+ D^-$ asymmetry
(see the previous figure).
Again adjusting only one free parameter we can roughly reproduce 
the main trend of the LHCb data. 
Please note that our approach predicts correct sign
of the asymmetry. In Ref.\cite{GMS2018} we showed also results
for $\sqrt{s}$ = 8 TeV.

\subsection{Asymmetries at low collision energies}

Our approach has distinct predictions at low energies.
Here we show our predictions for low energies.
Quite large asymmetries were found. 
As discussed in Ref.\cite{MS2018} detailed studies of the asymmetries 
at low energies are necessary to pin down or limit subleading
fragmentation.

\begin{figure}[!h]
\begin{center}
\includegraphics[width=0.5\textwidth]{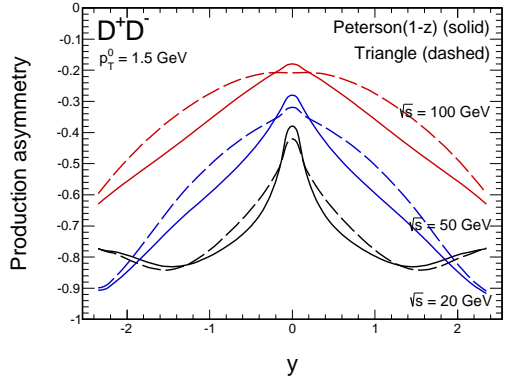}
  \caption{
\small $A_{D^{+}D^{-}}(y)$ production asymmetry in proton-proton
collisions for different center-of-mass energies $\sqrt{s}$.
}
\end{center}
\label{fig:assym_y_energy}
\end{figure}

\subsection{Charge-to-neutral $D$ meson ratio}

\begin{figure}[!h]
\begin{minipage}{0.42\textwidth}
  \centerline{\includegraphics[width=1.0\textwidth]{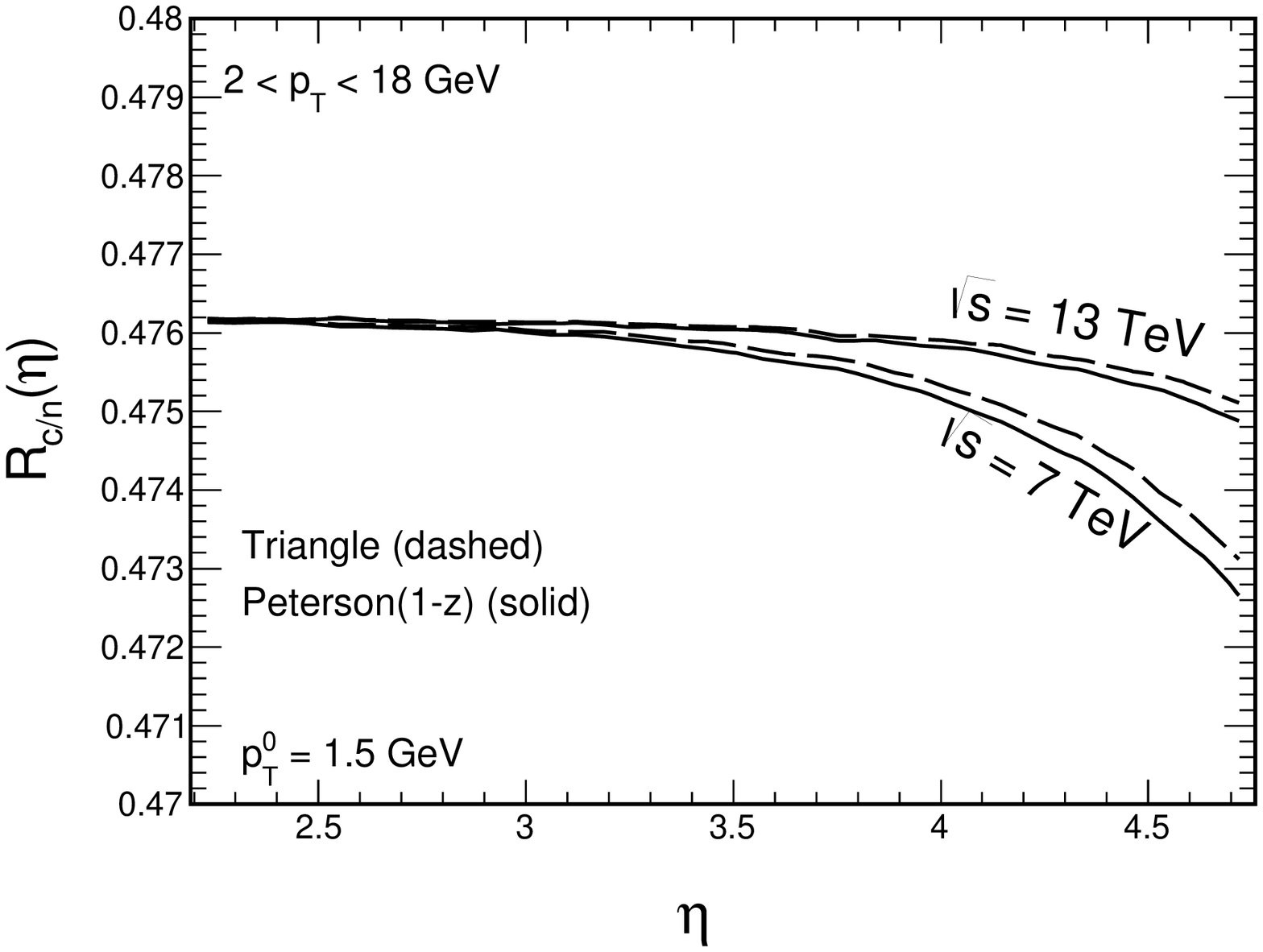}}
\end{minipage}
\begin{minipage}{0.42\textwidth}
  \centerline{\includegraphics[width=1.0\textwidth]{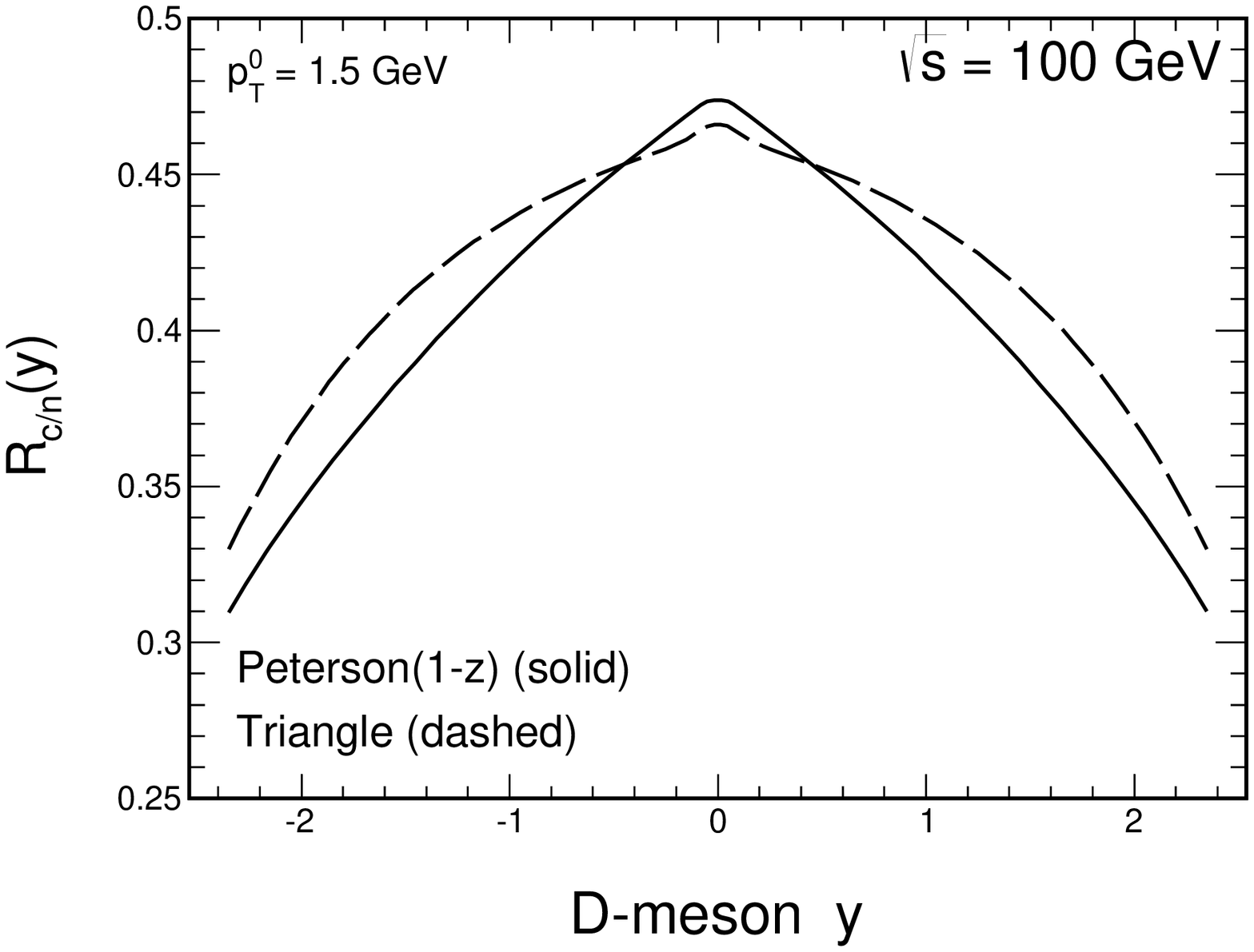}}
\end{minipage}
  \caption{
\small The $R_{c/n}$ ratio as a function of meson pseudorapidity for
$\sqrt{s}= 7$ and $13$ TeV for the LHCb kinematics (left panel) and
as a function of meson rapidity for $\sqrt{s}$ = 100 GeV 
in the full phase-space (right panel). 
Only quark-gluon components (diagrams) are included here in calculating
cross section for $q$ and $\bar q$ production.
}
\label{fig:R_cton}
\end{figure}

In Ref.\cite{MS2018} we discussed also the following ratio:
\begin{equation}
R_{c/n} \equiv \frac{D^+ + D^-}{D^0 + {\bar D}^0} \; .
\label{R_cton}
\end{equation}
In Fig.\ref{fig:R_cton} we show the ratio as a function of meson
rapidity for two different energies specified in the figure.
Evidently, when including subleading fragmentation, the ratio
depends on collision energy and rapidity. A test of such predictions
would be valuabale.

\subsection{$\nu_{\tau}$ neutrinos and ${\bar \nu}_{\tau}$
antineutrinos at IceCube}

\begin{figure}
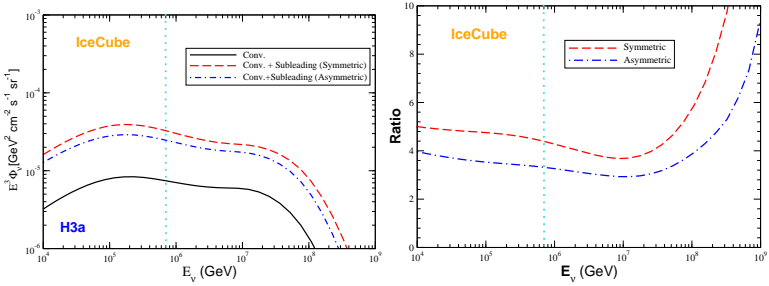

\begin{center}
\includegraphics[width=5cm]{tauneutrino.eps} 
\includegraphics[width=5cm]{ratiotauneutrino.eps}
\caption{
\small Our predictions for the flux of $\tau$ neutrinos (left panel)
and the suggested enhancement factor with respect to the traditional
$c {\bar c} \to D_s$ component (right panel).
}
\label{fig:flux_tau_neutrinos}
\end{center}
\end{figure}

In our recent analysis we showed how the flux of $\tau$ 
neutrinos/antineutrinos could be modified by the subleading 
$s/{\bar s} \to D_s$ fragmentation.
In Fig.\ref{fig:flux_tau_neutrinos} we show the conventional flux 
(due to $g g \to c \bar c$ fusion) and that of the subleading 
fragmentation (left panel) as well as the corresponding ratio (right panel).
The sizeable enhancement of the neutrino flux is not excluded in the moment.

\subsection{$\Lambda_c$ production}

\begin{figure}[!h]
\begin{minipage}{0.47\textwidth}
 \centerline{\includegraphics[width=0.9\textwidth]{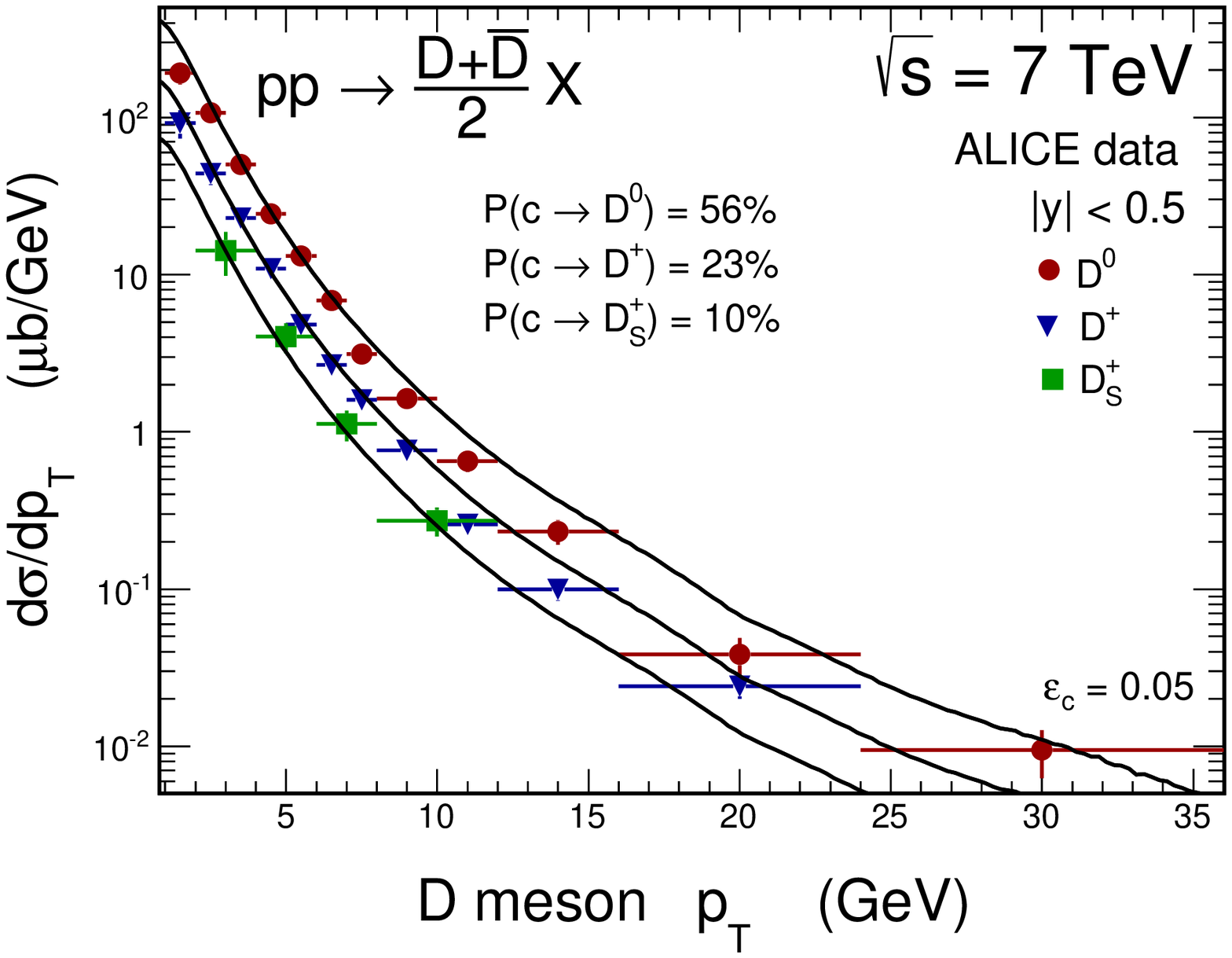}}
\end{minipage}
\begin{minipage}{0.47\textwidth}
 \centerline{\includegraphics[width=0.9\textwidth]{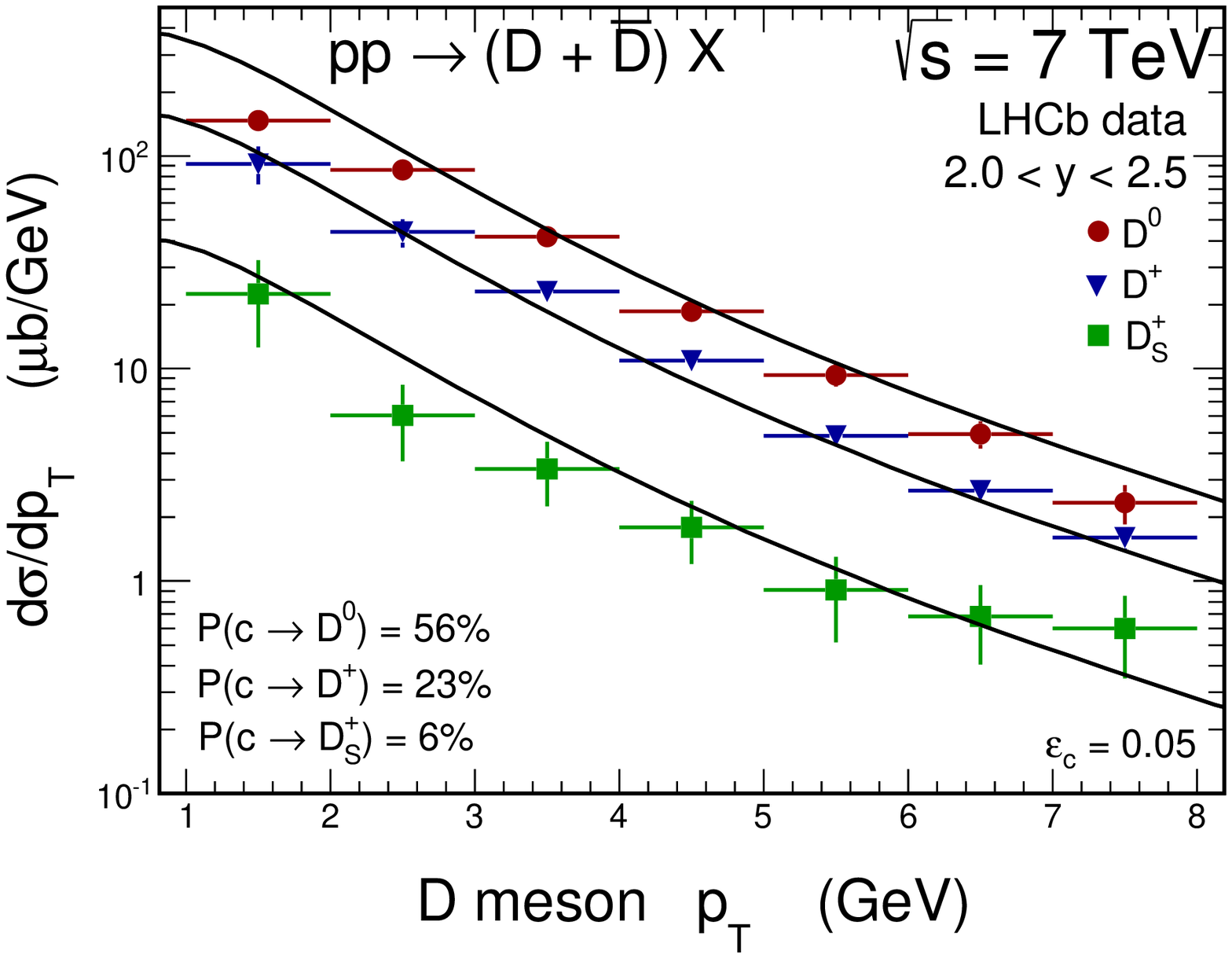}}
\end{minipage}
\caption{
\small Transverse momentum distribution of $D$ mesons for $\sqrt{s}$ = 7
TeV for ALICE (left panel) and LHCb (right panel). 
}
 \label{fig:dsig_dpt_Dmesons}
\end{figure}

In Fig.\ref{fig:dsig_dpt_Dmesons} we show our description of $D$ meson
transverse momenta. In this calculation $y_D = y_c$ was assumed.
This is a standard technical prescription for $c / {\bar c} \to D$ 
meson production in $pp$ collisions.

\begin{figure}[!htbp]
\begin{minipage}{0.47\textwidth}
 \centerline{\includegraphics[width=0.9\textwidth]{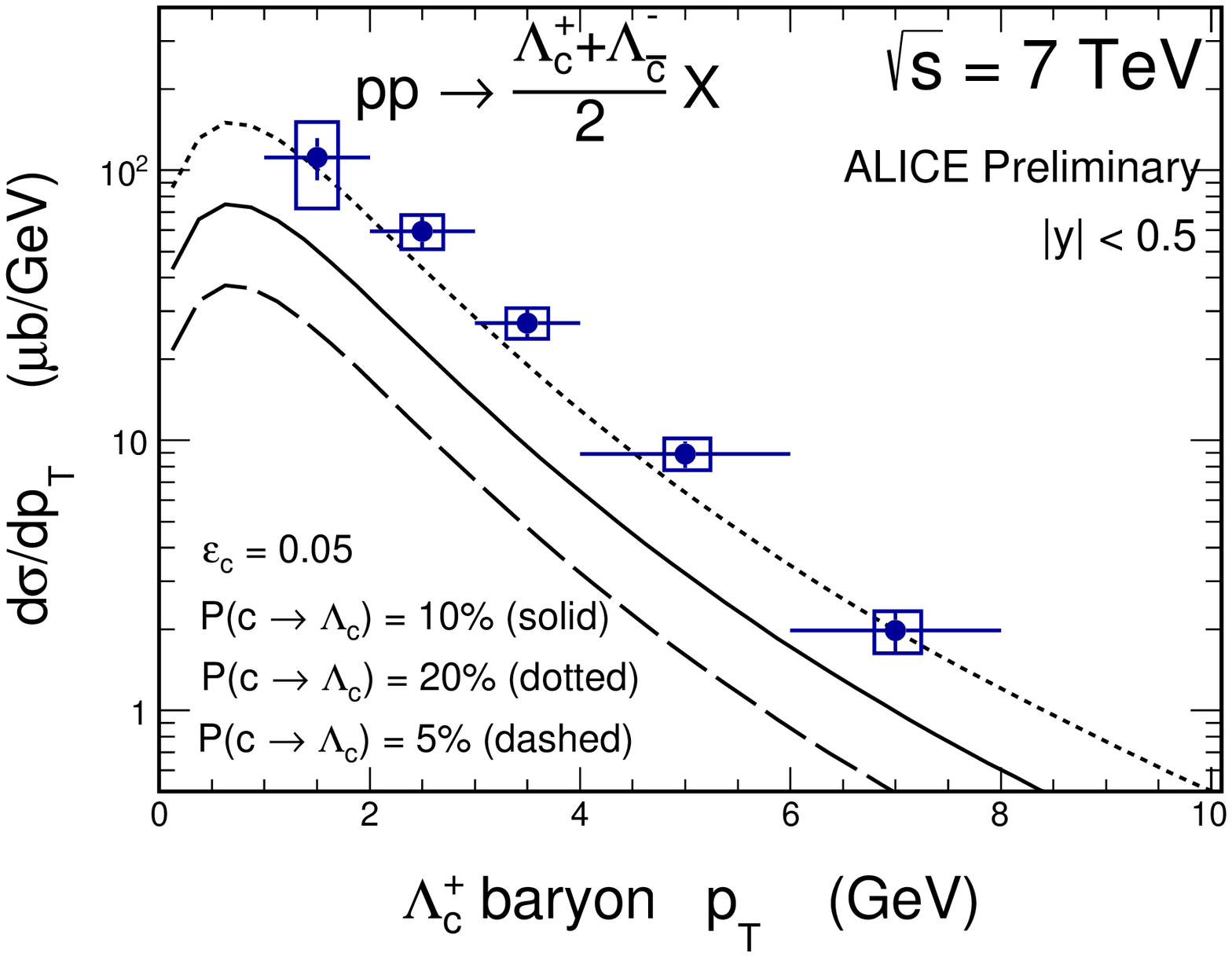}}
\end{minipage}
\begin{minipage}{0.47\textwidth}
 \centerline{\includegraphics[width=0.9\textwidth]{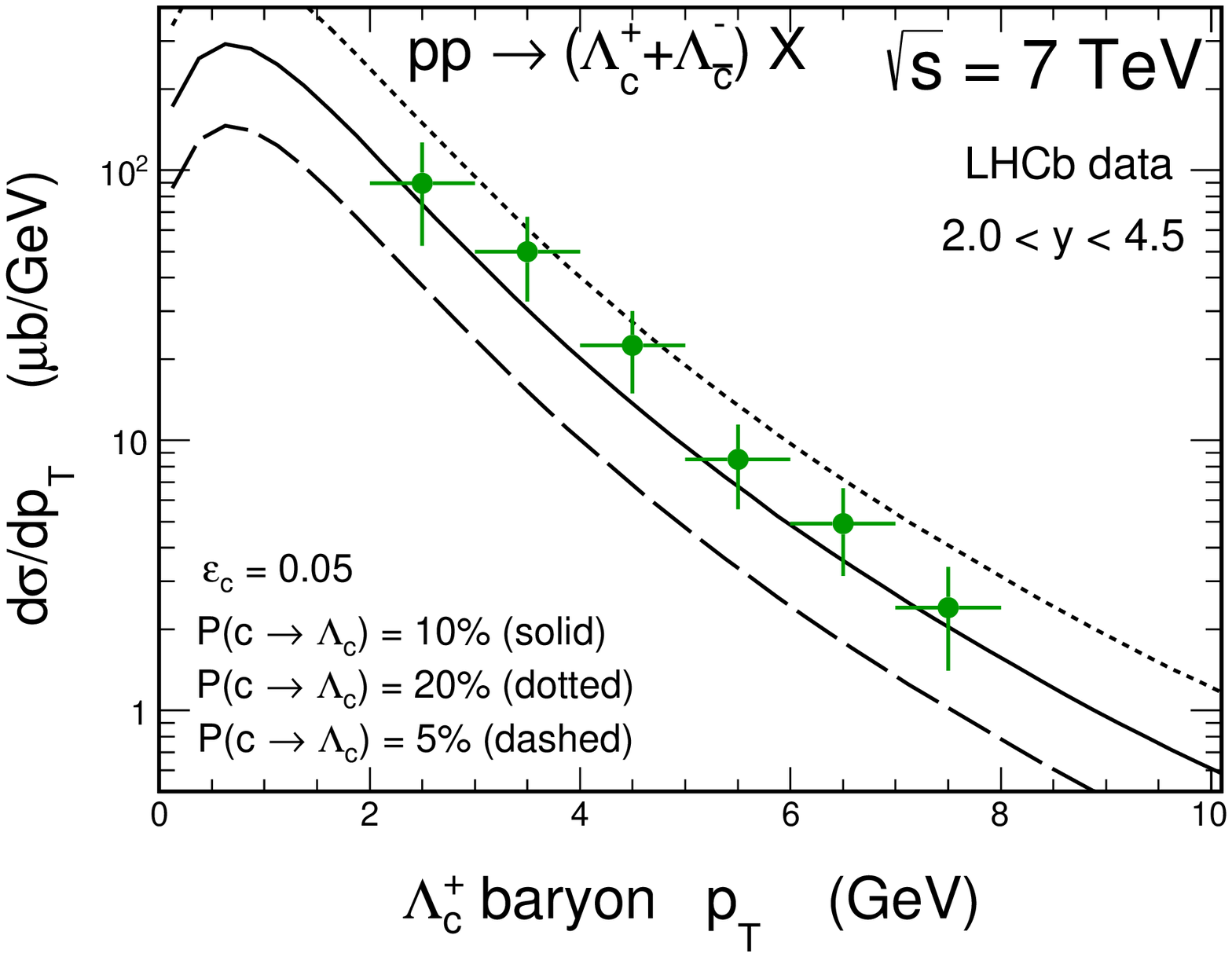}}
\end{minipage}
\caption{
\small Transverse momentum distribution of $\Lambda_c$ baryon 
for $\sqrt{s}$ = 7 TeV for ALICE (left panel) and LHCb (right panel).
}
\label{fig:dsig_dpt_LambdaC}
\end{figure}

In Fig.\ref{fig:dsig_dpt_LambdaC} we show similar results for
$\Lambda_c$ production. We have shown our results for different
$c / {\bar c} \to \Lambda_c$ transition probabilities.
Values of the transition probability smaller than 10 \% were
obtained from $e^+ e^-$ collisions. The new LHC data require
much larger transition probabilities. This is especially true for
the ALICE (midrapidity) data, where a value close to 20 \% is needed.
Does it signal a new mechanism?

\begin{figure}[!htbp]
\begin{minipage}{0.47\textwidth}
 \centerline{\includegraphics[width=0.9\textwidth]{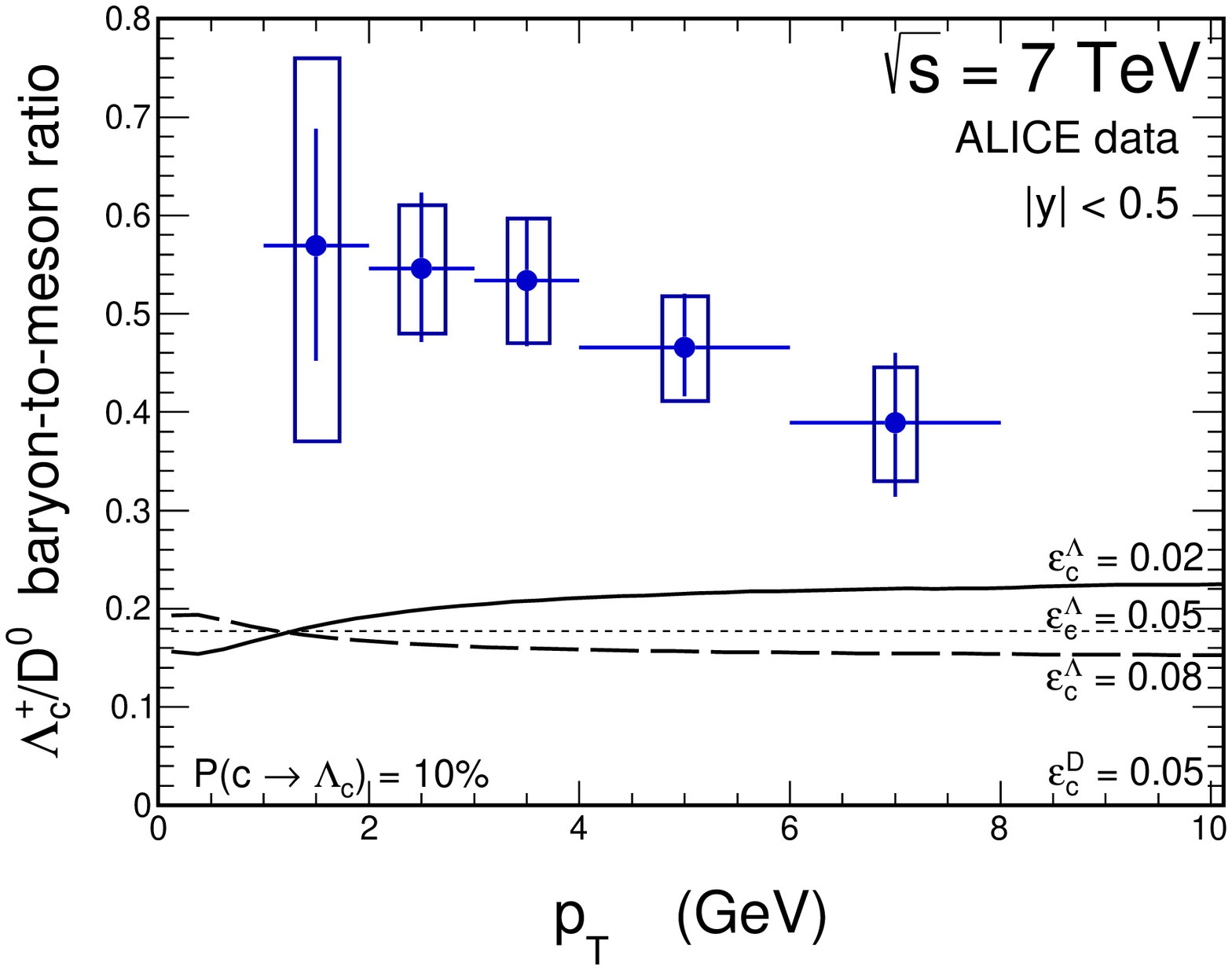}}
\end{minipage}
\begin{minipage}{0.47\textwidth}
 \centerline{\includegraphics[width=0.9\textwidth]{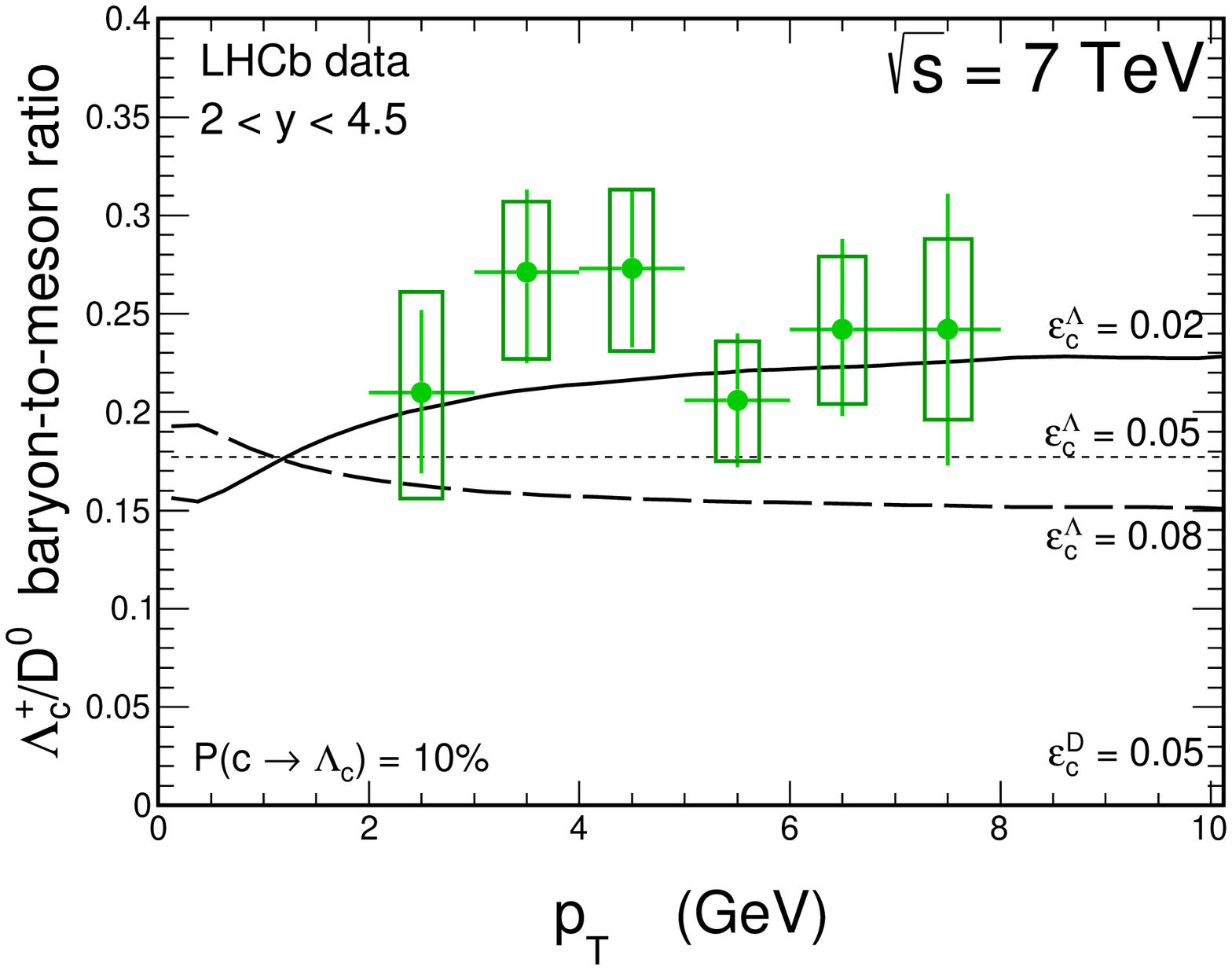}}
\end{minipage}
\caption{
\small Transverse momentum dependence of the
$\Lambda_c/D^0$ baryon-to-meson ratio for ALICE (left) and LHCb (right)
for different choices of the $\varepsilon_{c}^{\Lambda}$
parameter for $c \to \Lambda_c$ transition in the Peterson 
fragmentation function.
}
 \label{fig:ratio_pt_different_epsilons}
\end{figure}

In Fig.\ref{fig:ratio_pt_different_epsilons} we show the ratio of
cross section for $\Lambda_c$ to the cross section for $D^0$.
This once more shows a problem of independent-parton fragmentation
picture, especially at midrapidities.

In Ref.\cite{MS_Lambdac} we studied other options such as emissions
with the assumption $\eta_{\Lambda_c} = \eta_c$ (pseudorapidities) 
as well as possible feed down from highly excited charmed baryons. 
Some small improvements, especially for the ratio, are possible but 
the main disagreement with independent parton fragmentation picture stays.
Perhaps this could be explained in terms of a recombination model.
This requires further studies and modeling of such processes.

\section{Conclusions}

In one of our recent papers we demonstrated that the production of
high-energy neutrinos is related to very high $pp$ collision energies 
(even larger than at the LHC) and rather large $x_F$ 
(not accessible at the LHC).
Do we know mechanisms of $D$ meson production in these regions? 

Here we have presented and discussed briefly some results on asymmetry 
in the production of $D^+$ and $D^-$
\cite{MS2018} as well as $D_s^+ D_s^-$ mesons \cite{GMS2018} 
observed recently by the LHCb collaboration \cite{LHCb:2012fb,Aaij:2018afd}. 
Here we have discussed a scenario in which subleading
(unfavored) fragmentation $q/{\bar q} \to D^{\pm}$ is responsible 
for the asymmetry.
In the case of $D^+ D^-$ asymmetry it is quark-antiquark asymmetry
in the nucleon which is responsible for the effect. Adjusting the
corresponding quark/antiquark fragmentation probability we were able 
to describe the corresponding asymmetry.
This has dramatic consequences for low collision energies.
We predicted huge asymmetries for RHIC and even larger for lower energies.
We hope this will be verified in future by planned or possible
to perform experiments. It is not yet checked what are consequences
of the subleading fragmentation for high-energy neutrino production.

The asymmetry in the production of $D_s^+$ and $D_s^-$ mesons
is a bit more subtle. Here we have ${\bar s} \to D_s^+$ and $s \to D_s^-$
subleading fragmentations. The asymmetry of $D_s^+$ and $D_s^-$
production is possible provided there is $s(x) \ne {\bar s}(x)$.
Recently we have used one of the CTEQ parton distributions from
the fit which allows such a $s - \bar s$ asymmetry in longitudinal
momentum fraction. Our approach gives then correct sign of the asymmetry
and it was possible to find corresponding transition probability
to roughly describe the LHCb data.
This procedure was used to calculate flux of $\tau$ neutrinos
produced in the atmosphere. A significant enhancement was suggested.
There are first trials to identify
$\tau$ neutrinos with the help of IceCube aparatus \cite{Aartsen:2015dlt}.

Finally we have discussed production of $\Lambda_c$ baryons within
independent-parton fragmentation picture.
It was demonstrated that such a picture is insufficient to consistently
describe new LHC data. Especially for midrapidities (ALICE experiment) 
one observes a significant enhancement compared to the results
with corresponding fragmentation probabilities $c / {\bar c} \to \Lambda_c$
obtained from $e^+ e^-$ collisions as well as for lower proton-proton
collision energies. This strongly suggest a new mechanism.
Quark recombination is a good candidate.

{\bf Acknowledgments}

This study was partially supported by the Polish National Science Center
grant DEC-2014/15/B/ST2/02528 and by the Center for Innovation and
Transfer of Natural Sciences and Engineering Knowledge in Rzesz{\'o}w.



\end{document}